\documentclass[twocolumn,showpacs,prb,amsfonts,amsmath,amssymb,floatfix]{revtex4}
\usepackage{hhline}
\usepackage{mathrsfs}
\usepackage{graphicx}
\usepackage{dcolumn}
\usepackage{bm}% bold math

\input{epsf}

\newcounter{ichi}
\begin{document}
\preprint{cond-mat/0110430}
%------------------------------------------------
\title{Efficient and Accurate Linear Algebraic Methods for Large-scale Electronic Structure Calculations with 
Non-orthogonal Atomic Orbitals}
\author{H. Teng$^{1}$}
\altaffiliation[Present address: ]{Institute of Mechanics, Chinese Academy of Sciences, Beijing, China}
\author{T. Fujiwara$^{1,2}$}
\email[Corresponding author:\ ]{fujiwara@coral.t.u-tokyo.ac.jp}
\author{T. Hoshi$^{3,2}$}
\author{T. Sogabe$^{4,2}$}
\author{S.-L. Zhang$^{5,2}$}
\author{S. Yamamoto$^{6,2}$} 
\affiliation{{\rm (1)} Center for Research and Development of Higher Education, 
The University of Tokyo, Bunkyo-ku, Tokyo, 113-8656, Japan}
\affiliation{{\rm (2)} Core Research for Evolutional Science and Technology, 
Japan Science and Technology Agency (CREST-JST), Japan}
\affiliation{{\rm (3)} Department of Applied Mathematics and Physics, Tottori University, Tottori 680-8550, Japan}
\affiliation{{\rm (4)} School of Information Science and Technology, Aichi Prefecture University, Nagakute-cho, Aichi 480-1198, Japan}
\affiliation{{\rm (5)} Department of Computational Science and Engineering, Nagoya University, Chikusa-ku, Nagoya 464-8603, Japan}
\affiliation{{\rm (6)} School of Computer Science, Tokyo University of Technology, Katakura-machi, Hachioji, Tokyo 192-0982, Japan} 

%------------------------------------------------
\date{\today}
%%%%%%%%%%%%%%%%%%%%%%%%%%%%%%%%%%%%%%%%%%%%%%%%%
\begin{abstract}
The need for large-scale electronic structure calculations arises recently in the field of material physics
and efficient and accurate algebraic methods for large simultaneous linear equations become greatly important. 
We investigate the generalized shifted conjugate orthogonal conjugate gradient method, 
the generalized Lanczos method and the generalized Arnoldi method. 
They are the solver methods of large simultaneous linear equations of one-electron Schr\"odinger equation 
and maps the whole Hilbert space to a small subspace called the Krylov subspace. 
These methods are applied to systems of fcc Au with the NRL tight-binding Hamiltonian 
(Phys. Rev. B {\bf 63}, 195101 (2001)). 
We compare results by these methods and the exact calculation and 
show them equally accurate. 
The system size dependence of the CPU time is also discussed. 
The generalized Lanczos method and the generalized Arnoldi method are 
the most suitable for the large-scale molecular dynamics simulations 
from the view point of CPU time and memory size. 
\end{abstract} 
%------------------------------------------------
\maketitle 

%%%%%%%%%%%%%%%%%%%%%%%%%%%%%%%%%%%%%
%%%%%%    TEXT START    %%%%%% 
%%=================================== 
%%=================================== 
\section{Introduction}

In recent years, molecular dynamics (MD) simulations with electronic structure calculations 
in nano-scale structures have attracted much attention. 
One needs a large size of systems of several hundred thousands 
atoms with a few hundred pico-seconds (or more  longer time) process 
in order to investigate characteristics of nano-scale systems such as 
phenomena of competition between different physical principles or phenomena of the multi-physics, 
e.g. energy competition between the strain field and chemical bonds.~\cite{Si-fracture-1,Si-fracture-2,Au-multishell-1,Au-multishell-2}
Several requirements for large-scale MD simulation with electronic structure calculations 
are contradictory to each other, e.g. 
total energy accuracy {\it vs.} larger system size or longer physical time of processes.

There are several approaches for large-scale MD simulations;~\cite{Goedecker-1999}   
(a) the Fermi operator expansion,~\cite{Fermi-Op} 
(b) the divide-and-conquer method,~\cite{Devide-Conquer} and
(c) the minimization method (the density matrix minimization~\cite{DM-min} or the wavefunction minimization~\cite{Mauri-1993}).
Another classification may be the one according as the basis set of wavefunctions; 
(a) the plane wave basis set and switching between the real-space and $k$-space representation,~\cite{ONETEP} and 
(b) localized orbitals~\cite{SIESTA} or tight-binding basis set.~\cite{Ordejon-1998} 
Computation with ``massively parallel machine'' is also an important issue.

An important aspect is development of novel algebraic algorithm for extra-large scale systems. 
The most general and important algorithm may be the linear algebra 
solving simultaneous linear equations 
\begin{eqnarray}
    (z-H ){\bm x} ={\bm b} , 
\label{simul-1}
\end{eqnarray}
where $H$ is self-adjoint or real symmetric matrix, ${\bm b}$ is a given vector, 
$z= \varepsilon + i \eta$,  $\varepsilon$ is an energy parameter 
and $\eta$ is an infinitesimally small positive number, respectively. 
Solutions of Eq.(\ref{simul-1}) relate to the standard eigen-value problem $(\varepsilon-H ){\bm x}=0$. 
We developed the subspace diagonalization method and the shifted conjugate orthogonal conjugate gradient 
(COCG) method.~\cite{TAKAYAMA2004,TAKAYAMA2006,sQMR-SYMm,SOGABE2007} 
Then the methods were applied to 
the fracture propagation and surface formation in Si crystals 
with  the tight-binding Hamiltonian based on an orthogonal basis set.~\cite{Si-fracture-1,Si-fracture-2} 
On the other hand, since its Hamiltonian is described by the tight-binding Hamiltonian based on a non-orthogonal basis set, 
the problem of the formation of Au multishell helical nanowires was solved 
by the exact diagonalization method.~\cite{Au-multishell-1,Au-multishell-2}

Development of efficient linear algebraic methods has been, so far, mainly based on the orthogonal 
basis sets.~\cite{TAKAYAMA2004,TAKAYAMA2006,Lanczos,Haydock-1980,g-Eig-Bai} 
However, localized basis wavefunctions are generally non-orthogonal and 
it is much desirable to generalize the methods to the case of a non-orthogonal basis set.
The most popular strategy of the generalized eigen-value problem 
(represented by the non-orthogonal basis set) 
would be the transformation to the standard eigen-value problem.~\cite{g-Eig-Bai} 
Our target in the present paper is to solve simultaneous linear equations 
with self-adjoint or real symmetric matrix $S$; 
\begin{eqnarray}
    (zS-H ){\bm x} ={\bm b} ,
\label{simul-2}
\end{eqnarray}
which relates to the generalized eigen-value problem $(\varepsilon S-H){\bm x}=0$. 
We will investigate efficient methods of solving Eq.(\ref{simul-2}) 
with a complex energy variable $z$ when the matrix size of $H$ and $S$ is huge. 
Several algebraic algorithms will be discussed and directly applied to a tight-binding Hamiltonian 
based on non-orthogonal atomic orbitals in large-scale systems.

The structure of the present paper is as follows. 
In Section \ref{background}, the idea of non-orthogonal atomic orbitals and 
physical properties (e.g. the band energy, the local/partial density of states, 
numbers of occupied electron states, the chemical potential {\it et al.}) are summarized. 
Sections \ref{GsCOCG}, \ref{GLanczos} and \ref{GArnoldi} explain three different algorithms 
of large-scale linear equations, {\it i.e.} the generalized shifted conjugate orthogonal 
conjugate gradient method (GsCOCG), the generalized Lanczos method and the generalized Arnoldi method 
which generate the Krylov subspace from the whole Hilbert space.  
In these sections, numerical examples are presented by using the NRL tight-binding Hamiltonian.
The generalized Lanczos method becomes applicable to actual large systems with a high accuracy 
if one use  the modified Gram-Schmidt reorthogonalization to maintain the orthogonality of 
generated basis vectors. 
In Section \ref{Comparison}, we compare the CPU-times of each algorithm 
and discuss the applicability to large-scale electronic structure calculations and MD simulations. 
Section \ref{Conclusion} is conclusions. 
The examples without reorthogonalization in the generalized Lanczos method 
are shown and discussed in Appendix \ref{non-GS-method}.
Appendix \ref{EF-consistency} is devoted to discuss the consistency between the total energy and force. 

%%===================================
%%===================================  
\section{Theoretical background}\label{background}
%%-----------------------------------
\subsection{Non-orthogonal basis set and S-orthogonalization}

We define two sets of wavefunctions, $\{\phi_i({\bm r})\}$ and $\{\psi_\alpha({\bm r})\}$, 
where 
$\{\phi_i({\bm r})\}$ is the non-orthogonal (normalized) basis set 
(e.g. atomic orbitals and `$i$' denotes an atomic site and energy level), 
and $\{\psi_\alpha({\bm r})\}$ is the orthonormalized basis set. 
Then the overlap matrix $S_{ij}$ and the Hamiltonian matrix $H_{ij}$ are defined as 
\begin{eqnarray} 
S_{ij} &=& \langle \phi_i | \phi_j \rangle = \int\phi_i^* \phi_j d{\bm r}, \ \ \  S_{ii}=1 \ ,   
\label{eq(1.1)}\\
H_{ij} &=& \langle \phi_i |{\hat H}| \phi_j \rangle = \int\phi_i^* {\hat H} \phi_j d{\bm r}  \ \ \ ,
\label{eq(1.1-Hamiltonian)}
\end{eqnarray}
where ${\hat H}$ is the Hamiltonian operator.  
The orthonormal  basis set $\{\psi_\alpha({\bm r})\}$ can be expanded in terms of  $\{\phi_j\}$ as 
\begin{eqnarray}
\psi_\alpha ({\bm r}) &=& \sum_i \phi_i ({\bm r})w_i^{(\alpha)} 
\label{eq(1.2)} 
\end{eqnarray}
and the orthogonality relation is expressed as 
\begin{eqnarray}
 \langle \psi_\alpha | \psi_\beta \rangle 
 &=&    \sum_{ij} {w_i^{(\alpha)}}^*w_j^{(\beta)} S_{ij}  \label{eq(1.4)}     \\
 &\equiv& ({\bm w}^{(\alpha)}, {\bm w}^{(\beta)})_S   
  = \delta_{\alpha \beta}  \ \ ,                         \label{eq(1.4p)} 
\end{eqnarray}
where ${\bm w}^{(\alpha)}=(w_1^{(\alpha)},w_2^{(\alpha)},\cdots\cdots)^{\rm t}$.  
We call the representation $({\bm w}^{(\alpha)}, {\bm w}^{(\beta)})_S$ the ``S-product" 
and the relation Eq.~(\ref{eq(1.4p)}) the ``S-orthogonalization" of basis vectors ${\bm w}^{(\alpha)}$.

When $\psi_\alpha ({\bm r})$ satisfies the Schr\"odinger equation  
\begin{eqnarray}
{\hat H} \psi_\alpha ({\bm r})= \varepsilon_\alpha \psi_\alpha ({\bm r})  ,
\label{eq(1.5)}
\end{eqnarray}
coefficients $\{w_i^{(\alpha)}\}$ should be elements of 
an eigen-vector of a simultaneous linear equation in the $\phi$-representation;  
\begin{eqnarray}
\sum_i  H_{ji} w_i^{(\alpha)}=\varepsilon_\alpha \sum_iS_{ji} w_i^{(\alpha)}
\label{eq(1.8)}
\end{eqnarray}
or, in matrix-vector form,  
\begin{eqnarray}
  H {\bm w}^{(\alpha)}=\varepsilon_\alpha S {\bm w}^{(\alpha)}.
\label{eq(1.9)}
\end{eqnarray}
Matrices $H=(H_{ij})$ and $S=(S_{ij})$ are self-adjoint in $\phi$-representation.

%%=================================== 
\subsection{Green's function and local/partial density of states represented by the non-orthogonal basis set}

The Green's operator ${\hat G}$ is defined as
\begin{eqnarray}
    {\hat G}(z)= \{(\varepsilon+i\eta){\hat 1}-{\hat H}\}^{-1} , 
\label{eq(3.3}
\end{eqnarray}
where ${\hat 1}$ is the identity operator and $z=\varepsilon+i\eta$.  
Elements of the Green's function matrix can, then, be written as 
\begin{eqnarray}
G_{ij}(z) &=& \langle \phi_i|{\hat G}(z)|\phi_j\rangle =\{S(zS-H)^{-1}S\}_{ij}  \label{eq(3.8-p)} \\
          &=&  \sum_{k, l} S_{ik} \Big\{\sum_\alpha {w_k^\alpha}^* \frac{1}{z-\varepsilon_\alpha} w_l^\alpha \Big\}S_{l j} .
\label{eq(3.8)}
\end{eqnarray}
The local (partial) densities of states is expressed in the $\phi$-representation as follows:
\begin{eqnarray}
 D_{ij}(\varepsilon ) &=&  -\frac{1}{\pi}{\rm Im} (G(z)S^{-1})_{ij} . 
\label{eq(3.18)} 
\end{eqnarray}
The normalization of the Green's functions and the local/partial density of states (DOS) is then 
\begin{eqnarray}
&& \Big(-\frac{1}{\pi}\Big)\int_{-\infty}^\infty d \varepsilon {\rm Im}G_{ij}(z)
   = \langle \phi_i|\phi_j\rangle =S_{ij}                      ,   \label{eq(3.9)} \\
&& \int _{-\infty}^\infty d\varepsilon D_{ii}(\varepsilon ) =1 .   \label{eq(3.20)} 
\end{eqnarray}

%%=================================== 
\subsection{Total band energy and Green's function}

%%-----------------------------------
\subsubsection{Density matrix and energy density matrix}

In the simulation process, the density matrix $\rho_{ij}$ and the energy density matrix 
$\pi_{ij}$ appear repeatedly in the calculation of the Mulliken charge, the total energy 
and forces,~\cite{1998-Elstner}
whose definition may be  
\begin{eqnarray}
\rho_{ij}
 &=& \Big(-\frac{1}{\pi}\Big) {\rm Im} \int d\varepsilon 
     \sum_\alpha f(\varepsilon_\alpha) \frac{w_i^{(\alpha)*} w_j^{(\alpha)}}{z-\varepsilon_\alpha}   \label{eq(4.83)} \\
 &=& \sum_{\alpha} f(\varepsilon_\alpha) w_i^{(\alpha)*}w_j^{(\alpha)}  ,                            \label{eq(4.84)} \\
\pi_{ij} 
 &=& \Big(-\frac{1}{\pi}\Big) {\rm Im} \int d\varepsilon \varepsilon 
     \sum_\alpha f(\varepsilon_\alpha) \frac{w_i^{(\alpha)*} w_j^{(\alpha)}}{z-\varepsilon_\alpha}   \label{eq(4.86)} \\
 &=& \sum_{\alpha} f(\varepsilon_\alpha)  \varepsilon_\alpha w_i^{(\alpha)*}w_j^{(\alpha)}  ,        \label{eq(4.87)}
\end{eqnarray}
where $f(\varepsilon)$ is the Fermi-Dirac function  $f(\varepsilon)=\{1+\exp((\varepsilon-\mu)/\tau)\}^{-1}$, 
where $\mu$ and $\tau$ are the chemical potential and temperature.

%%-----------------------------------
\subsubsection{Physical property}\label{ChemPot-BandEne}

The chemical potential $\mu$ should be determined 
by the equation for the total electron number $N_{\rm tot}$:
\begin{eqnarray}
N_{\rm tot} 
&=& 2\sum_i \int d\varepsilon f(\varepsilon) D_{ii}(\varepsilon)
\label{eq(4.79)}\\
&=& \sum_{ij\alpha} S_{ij} \Big(-\frac{2}{\pi}\Big)   
    {\rm Im}\int d\varepsilon f(\varepsilon_\alpha)  \frac{w_j^{(\alpha)*} w_i^{(\alpha)}}{z-\varepsilon_\alpha}
\label{eq(4.80)}\\
&=& 2 \sum_{ij} S_{ij} \rho_{ji}  ,
\label{eq(4.81)} 
\end{eqnarray}
where a factor ``2'' is the spin degeneracy.

The total band energy of the system is given as
\begin{eqnarray}
E_{\rm tot}
&=& 2 \sum_{\alpha}^{\rm occ} \varepsilon_\alpha  = 2 \sum_{\alpha} \varepsilon_\alpha f(\varepsilon_\alpha )
\label{eq(3.1)} \\
&=& -\frac{2}{\pi}{\rm Im} \sum_i \int d\varepsilon \varepsilon f(\varepsilon) (G(z)S^{-1})_{ii} ,
\label{eq(3.16)}
\end{eqnarray}
where the summation $\sum_{\alpha}^{\rm occ}$ runs over the occupied states.
This equation can be expressed by the density of states, the density matrix or the energy density matrix as
\begin{eqnarray}
E_{\rm tot} 
&=& 2\sum_i \int d\varepsilon \varepsilon f(\varepsilon) D_{ii}(\varepsilon)                    
\label{eq(4.88)}\\
&=& 2 \sum_{ij} \rho_{ij} H_{ji}     
\label{eq(4.89)}\\
&=& 2 \sum_{ij} S_{ij} \pi_{ji} .    
\label{eq(4.90)}
\end{eqnarray}

Moreover, any physical property can be expressed by using 
the density matrix as 
\begin{eqnarray}
\langle X \rangle  
&=& \Big(-\frac{2}{\pi}\Big) \int d\varepsilon f(\varepsilon) \sum_{ij} 
     X_{ij}{\rm Im} (S^{-1}G(z)S^{-1})_{ji} \nonumber \\
&=& 2\sum_{ij}X_{ij}\rho_{ji}. 
\label{eq(GeneralPhyPro)} 
\end{eqnarray}
The expressions Eqs.~(\ref{eq(4.89)}) and (\ref{eq(4.90)}) and also (\ref{eq(GeneralPhyPro)}) 
are satisfied not only in the whole Hilbert space but also in the mapped subspace 
in which we construct approximate eigen-states.

Now we have obtained three different expressions 
Eqs.(\ref{eq(4.79)})$\sim$(\ref{eq(4.81)}) for $N_{\rm tot}$ and 
Eqs.(\ref{eq(4.88)})$\sim$(\ref{eq(4.90)}) for $E_{\rm tot}$. 
These expressions normally give different values, 
because we usually use finite values of the energy interval, $\eta$ and 
approximate eigen-states in the mapped subspace.  
Fortunately, if the formula $\sum_{ij} \rho_{ij} H_{ji} = \sum_{ij} S_{ij} \pi_{ji}$ 
is satisfied, the consistency between the total band energy and the force can be kept 
as shown in Appendix \ref{EF-consistency}. 

%%=================================== 
%%=================================== 
\section{Generalized shifted COCG method}\label{GsCOCG}
We developed the shifted COCG method 
for large-scale linear equations (\ref{simul-1}).~\cite{TAKAYAMA2006,sQMR-SYMm,fujiwara-2008}
It was shown that the convergence behavior can be monitored by observing the behavior of the ``residual norm''. 
The shifted COCG method is generalized for Eq.(\ref{simul-2}) in this section. 

%=================================== 
\subsection{Definition of the problem}

The eigen-value problem of stationary Schr\"odinger equation is equivalent to 
the scattering problem;
\begin{eqnarray}
    (z{\hat 1}-{\hat H} ) \psi({\bm r}) = \chi({\bm r}) ,
\label{eq(B.1)}
\end{eqnarray}
where $z=\varepsilon+i\eta$ and $\varepsilon$ is an energy parameter of incident waves. 
The wavefunction $\psi({\bm r})$ is expanded by the set of non-orthogonal atomic orbitals $\{\phi_j\}$;
\begin{eqnarray}
    \psi({\bm r}) =\sum_j \phi_j({\bm r}) x_j(z)  .
\label{eq(B.2)}
\end{eqnarray}
Substituting Eq.~(\ref{eq(B.2)}) into Eq.~(\ref{eq(B.1)}), 
one obtains generalized linear equations 
\begin{eqnarray}
     (zS-H){\bm x}{(z)} = {\bm b} ,                    \label{eq(B.5)} 
\end{eqnarray}   
where the $j$-th component of the vector ${\bm b}$ is $b_j=\langle \phi_j|\chi \rangle$. 
The solution of the linear equation  ${\bm x}(z)$ is then 
\begin{eqnarray}
 {\bm x}(z)= (zS-H)^{-1} {\bm b}=S^{-1}G(z)S^{-1}{\bm b}
\label{eq(B.9)}
\end{eqnarray}   
with a help of Eq.~(\ref{eq(3.8-p)}). 
By setting a vector ${\bm b}$ as 
\begin{eqnarray}
 {\bm b}={\bm e}_{j} =(0,0,\cdots,0,\underbrace{1}_{j},0,0 \cdots )^{\rm t},
\label{eq(B.13)}
\end{eqnarray} 
we can get the corresponding solution ${\bm x}^j(z)$ as 
\begin{eqnarray}
 {\bm x}^j(z)= S^{-1}G(z)S^{-1}{\bm e}_j   .
\label{eq(B.9-1)}
\end{eqnarray}  
The product of ${\bm e}_i$ and ${\bm x}^j$ (not S-product), the $i$-th element of a vector ${\bm x}^j$, 
is identical just to the energy-component of the density matrix $\rho_{ij}(\varepsilon)$; 
\begin{eqnarray}
  \rho_{ij}(\varepsilon)
 &=& -\frac{1}{\pi}{\rm Im}\Big[{\bm e}^{\rm t}_i \cdot {\bm x}^j(\varepsilon+i\eta )\Big], 
\label{eq(B.18)}
\end{eqnarray}   
which relates to the local DOS as 
\begin{eqnarray}
  D_{ii}(\varepsilon) 
&=& \sum_k S_{ik}\rho_{ki}(\varepsilon)  .
\label{eq(B.20)}
\end{eqnarray}   
Then the density matrix and the energy density matrix are given by the integrations 
of  $\rho_{ij}(\varepsilon)$ as
\begin{eqnarray}
  \rho_{ij}
&=& \int d\varepsilon f(\varepsilon) \rho_{ij}(\varepsilon )              , \label{eq(B.19)} \\
  \pi_{ij}
&=& \int d\varepsilon \varepsilon f(\varepsilon) \rho_{ij}(\varepsilon )  . \label{eq(B.19-p)} 
\end{eqnarray}
It should be noticed here that there is no quantities of 
eigen-energies in the Krylov subspace and, we should use the calculation procedure 
through $\rho_{ij}(\varepsilon )$ rather than the calculation of Eqs.(\ref{eq(4.84)}) and (\ref{eq(4.87)}).  
Their resultant values depend on the interval of energy mesh-points for the energy integration 
and a fictitious finite value of $\eta$.

%=================================== 
\subsection{Generalized shifted conjugate orthogonal conjugate gradient (GsCOCG) method}

For non-orthogonal basis set, we can generalize the shifted COCG procedure, 
named the generalized shifted COCG (GsCOCG) method.~\cite{GsCOCG}~   
The linear equations of the `seed' energy $\sigma_s$ and 
the `shift' energy $\sigma$, respectively, are written down as 
\begin{eqnarray}
&&(S^{-1} A +\sigma_s 1){\bm x}       = S^{-1}{\bm b} \label{seed-E}  , \label{eq(B.21)}\\   
&&(S^{-1} A +\sigma   1){\bm x}^{(i)} = S^{-1}{\bm b} \label{shift-E} , \label{eq(B.22)}
\end{eqnarray}   
where the matrix $A$ is defined as  
\begin{eqnarray}
A=z_{\rm ref}S-H 
\label{defA}
\end{eqnarray}
with an arbitrary reference energy $z_{\rm ref}=\varepsilon_{\rm ref}+i\eta$, 
$1$ is the unit matrix and ${\bm b}={\bm e}_j$.
The seed energy and the `shift energy' are given as 
$\varepsilon_s=\varepsilon_{\rm ref}+\sigma_s$ and $\varepsilon=\varepsilon_{\rm ref}+\sigma$.

Following the procedure of the shifted COCG method,~\cite{sQMR-SYMm,SOGABE2007}
we try to find iterative $n$-th solutions ${\bm x}_n$ in the Krylov subspace defined  
\begin{eqnarray}
&&K_n(S^{-1}A+\sigma_s I, S^{-1}{\bm b}) \nonumber \\
&&= {\rm Span}\{S^{-1}{\bm b},S^{-1}AS^{-1}{\bm b}, (S^{-1}A)^2S^{-1}{\bm b}, \nonumber \\
&& \ \ \ \ \cdots, (S^{-1}A)^nS^{-1}{\bm b}\}  .
\label{Krylov-subS}
\end{eqnarray}
This yields the residual vector ${\bm r}^\prime_n=S^{-1}{\bm b}-(S^{-1}A+\sigma_S 1){\bm x}_n$ 
to be~\cite{GsCOCG}
\begin{eqnarray}
{\bm r}^\prime_n \perp K_n((S^{-1}A+\sigma_S 1)^\dagger, {\bm b}^*) ,
\end{eqnarray} 
where $B^\dagger$ is the Hermitian conjugate matrix of $B$ and ${\bm b}^*$ is the complex conjugate vector of ${\bm b}$.  
The actual algorithms may be as follows. 
Under the initial conditions
\begin{eqnarray}
 &&{\bm x}_0={\bm p}_{-1}={\bm 0},          \label{eq(B.23)} \\ 
 &&{\bm r}_0={\bm b},                       \label{eq(B.24)} \\
 && \alpha_{-1}=1, \ \ \  \beta_{-1}=0,     \label{eq(B.25)}
\end{eqnarray}
and a definition ${\bm r}_0^\prime =S^{-1}{\bm r}_0$,   
we evaluate the following equations for the `seed' energy $\sigma_s$ iteratively for $n=0,1,2,\cdots$:   
\begin{eqnarray}
 {\bm p}_n   &=& {\bm r}_{n}^\prime   + \beta_{n-1} {\bm p}_{n-1}  , 
\nonumber \\ 
 \alpha_{n}  &=& \frac{({\bm r}_{n}^\prime, {\bm r}_{n}^\prime)_S}{({\bm p}_{n}, S^{-1}(A +\sigma_sS) {\bm p}_{n})_S}  , 
\nonumber \\ 
{\bm x}_{n+1}&=& {\bm x}_{n} + \alpha_{n} {\bm p}_{n} ,  
\label{G-COCG} \\ 
{\bm r}_{n+1}&=& {\bm r}_{n} - \alpha_{n} (A + \sigma_sS) {\bm p}_{n} , 
\nonumber \\ 
{\bm r}_{n+1}^\prime &=& S^{-1}{\bm r}_{n+1} , 
\nonumber \\ 
 \beta_{n}  &=& \frac{({\bm r}_{n+1}^\prime,{\bm r}_{n+1}^\prime)_S}{({\bm r}_{n}^\prime,{\bm r}_{n}^\prime)_S}  . 
\nonumber     
\end{eqnarray}
Important point here is our use of ${\bm r}_n^\prime = S^{-1}{\bm r}_n$. 
In actual procedure, we employ a form ${\bm r}_n=S{\bm r}_n^\prime$ at each iteration step by CG method. 
Since the overlap matrix $S$ is real symmetric positive definite and sparse, 
the convergence of CG iteration can be fast.

The basic theorem of the Krylov subspace is the invariance
of the subspace under an energy shift $\sigma$.   
The other very basic theorem is 
the {\it collinear residual}~\cite{FROMMER2003}
\begin{equation}
	{\bm r}^\sigma_{n}=\frac{1}{\pi^{\sigma}_{n}}{\bm r}_{n}.
\label{eq(B.37)}
\end{equation}
Owing to these theorems, once we solve the set of equations for the `seed' energy $\sigma_s$, 
we can obtain the results for any shift energy $\sigma$ only by scalar multiplications. 
The recurrence equations for shift energies  are given 
(all the quantities are denoted by the superscript $\sigma$),  
with initial values $\pi^\sigma_{-1}=\pi^\sigma_0=1$, as follows; 
\begin{eqnarray}
	\pi^{\sigma}_{n+1}  &=& \Big\{1+ \alpha_n (\sigma-\sigma_s) \Big\} \pi^\sigma_n
                            + \frac{\beta_{n-1}}{\alpha_{n-1}}\alpha_{n}( \pi^\sigma_{n}- \pi^\sigma_{n-1})   \nonumber \\
\label{eq(B.32)}    
\end{eqnarray}
and  
\begin{eqnarray}
      {\bm x}^{\sigma}_{n+1}&=& {\bm x}^\sigma_{n} + \alpha^\sigma_{n} {\bm p}^\sigma_{n}   \label{eq(B.36)}  
\end{eqnarray}
with 
\begin{eqnarray*}
	\alpha^{\sigma}_n   &=&\frac{\pi^\sigma_n}{\pi^\sigma_{n+1}}\alpha_n                                      ,
\label{eq(B.33)} \\
	\beta^{\sigma}_{n-1}&=&\Big(\frac{\pi^\sigma_{n-1}}{\pi^\sigma_{n}}\Big)^2 \beta_{n-1}                            ,
\label{eq(B.34)}  \\
      {\bm p}^{\sigma}_{n}  &=& \frac{1}{\pi^\sigma_n}{\bm r}_{n}^\prime + \beta^\sigma_{n-1} {\bm p}^\sigma_{n-1} .
\label{eq(B.35)}     \\ 
\end{eqnarray*}

\begin{figure}[h]
\begin{center}
\includegraphics[width=8cm]{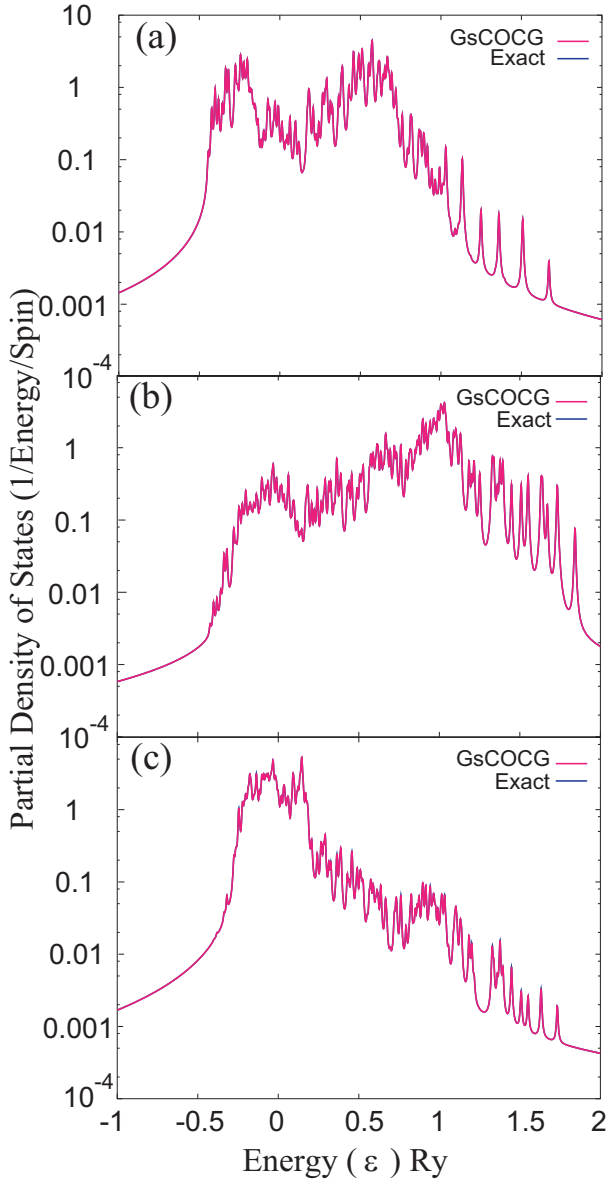}
\caption{(Color on line) Partial density of states for a system of Au 864 atoms by NRL 
tight-binding Hamiltonian,~\cite{tb-NRL} 
normalized to unity.  
(a) s-orbitals, (b) p-orbitals and (c) d-orbitals. 
Comparison is for GsCOCG and the exact calculation, which are almost identical to each other. 
Parameters in GsCOCG calculations are $\eta=10^{-3}$Ry, $\tau=5\eta$. 
The energy interval of mesh-points is $10^{-4}$Ry. 
See Figs.~\ref{fig_e-GL_Reorthogonalization}, \ref{fig_Arnoldi-Lanczos} and \ref{fig_e-GL_NoReorthogonalization}
for comparison.
}
\label{fig_GsCOCG-DOS}
\end{center}
\end{figure}

Partial densities of states are shown in Fig.~\ref{fig_GsCOCG-DOS}
for a system of Au 864 atoms by NRL tight-binding Hamiltonian,~\cite{tb-NRL} 
in comparison with those by the exact calculations. 
In order to see the behavior of the peak positions and the tail of the peaks, 
the figures are drawn in the logarithmic scale. 
Two lines of GsCOCG and the exact calculation overlap each other almost completely and  
one can recognize an excellent agreement between the two different calculations.

%- - - - - - - - - - Figure 1 GsCOCG- - - - - - - - - -%
%------------------------------------------------------%

%------------------------------------------------------------------------
\subsection{Residual norm and convergence behavior}

%- - - - - - - - - - Figure 2 GsCOCG- - - - - - - - - -%
\begin{figure}[h]
\begin{center}
\includegraphics[width=\linewidth]{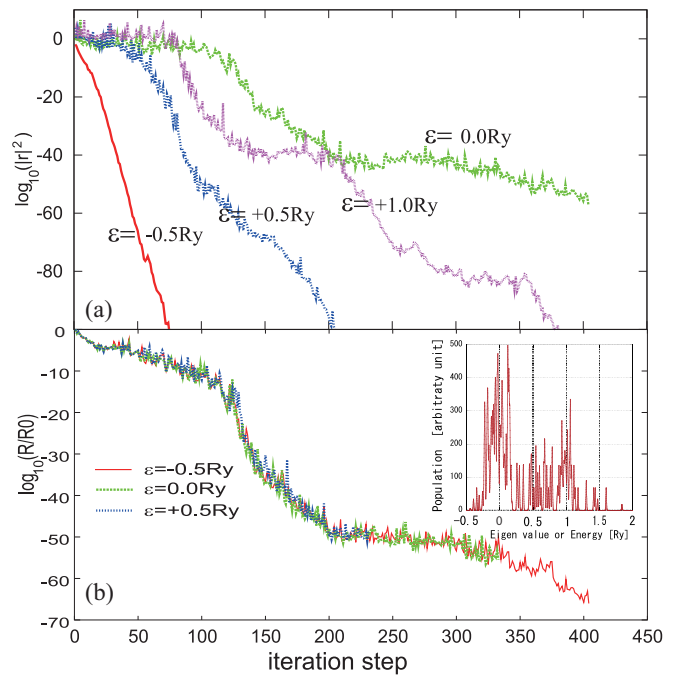}
\caption{(Color on line) Convergence behavior of residual norms for a system of Au 256 atoms by NRL 
tight-binding Hamiltonian.~\cite{tb-NRL} 
Spectrum extends between $-0.5$~Ry to $1.5$~Ry. 
The inset in (b) show the total density of states $D(\varepsilon)=\sum_i D_{ii}(\varepsilon)$, 
where we use a finite imaginary number in the energy and the profile is of dense spiky peaks.
(a) Residual norm $\parallel \bm{r}_n^{(s,j)} \parallel$ at several energy points 
$\varepsilon= -0.5,\ 0.0,\ 0.5,\ 1.0$~Ry for s-orbital.   
(b) Average residual norms $R_n^{(j)}$ with different three seed energies ($-0.5,\ 0.0,\ 0.5$Ry) 
for s-orbital and they all overlap with each other.
}
\label{fig_GsCOCG-ResidualNorm}
\end{center}
\end{figure}
%------------------------------------------------------%

The useful characteristic property of GsCOCG method is 
the capability of monitoring the norm of residual vectors.~\cite{TAKAYAMA2006}
The residual vectors for the seed and shift equations 
with an energy $\varepsilon_k$ (with ${\bm b}={\bm e}_j$ and 
$\sigma_k=\varepsilon_k-\varepsilon_{\rm ref}$) are $\bm{r}_n^{(s,j)}$ and $ \bm{r}_n^{(k,j)}$, 
respectively, and the \lq mapped' residual vectors 
for the seed and shift equations $ \bm{r}_n^{\prime(s,j)}=S^{-1}\bm{r}_n^{(s,j)}$ and 
$\bm{r}_n^{\prime(k,j)}=S^{-1}\bm{r}_n^{(k,j)}$. 
We usually need only elements of the density matrix among near-sited orbital pairs 
connected by non-zero elements of the Hamiltonian or overlap matrices 
and the convergence monitoring is necessary for these components.~\cite{TAKAYAMA2006} 
Therefore, in order to monitor the convergence behavior, we adopt the ``residual norm'' 
defined as 
\begin{eqnarray}
\parallel {\bm r}_n^{\prime(s/k,j)} \parallel^2 \equiv \sum_i^{H_{ij} \ne 0} | \bm{e}_i^{\rm t}\cdot {\bm r}_n^{\prime(s/k,j)} |^2 . 
\end{eqnarray}
Furthermore, since the residual norm is different among different energy points, 
the average quantity (``average residual norm'') should be defined as~\cite{TAKAYAMA2006}
\begin{eqnarray}
R_n^{(j)} 
&\equiv& \frac{1}{N_{\rm ene}} \sum_k^{N_{\rm ene}} \parallel \bm{r}_n^{\prime(k,j)} \parallel^2  \nonumber \\
&=     & \parallel \bm{r}_n^{\prime(s,j)} \parallel^2 \frac{1}{N_{\rm ene}} \sum_k^{N_{\rm ene}} \frac{1}{|\pi_k|^2} ,
\label{EQ-RES-NORM-AVE}
\end{eqnarray}
where $N_{\rm ene}$ is the number of energy points.

The convergence behavior of the residual norm for different seed energies 
is shown in Fig.~\ref{fig_GsCOCG-ResidualNorm}a. The convergence at the energy of the low DOS 
is very fast, because the eigen-state can be constructed by a small number of basis states. 
The convergence of the averaged norm is shown in Fig.~\ref{fig_GsCOCG-ResidualNorm}b, 
which confirms numerically the fact that the average residual norm (and all the physical quantities) 
does not depend sensitively on the choice of a seed energy. 

%- - - - - - - - - - Figure 3 GsCOCG- - - - - - - - - -%
\begin{figure}[h]
\begin{center}
\includegraphics[width=\linewidth]{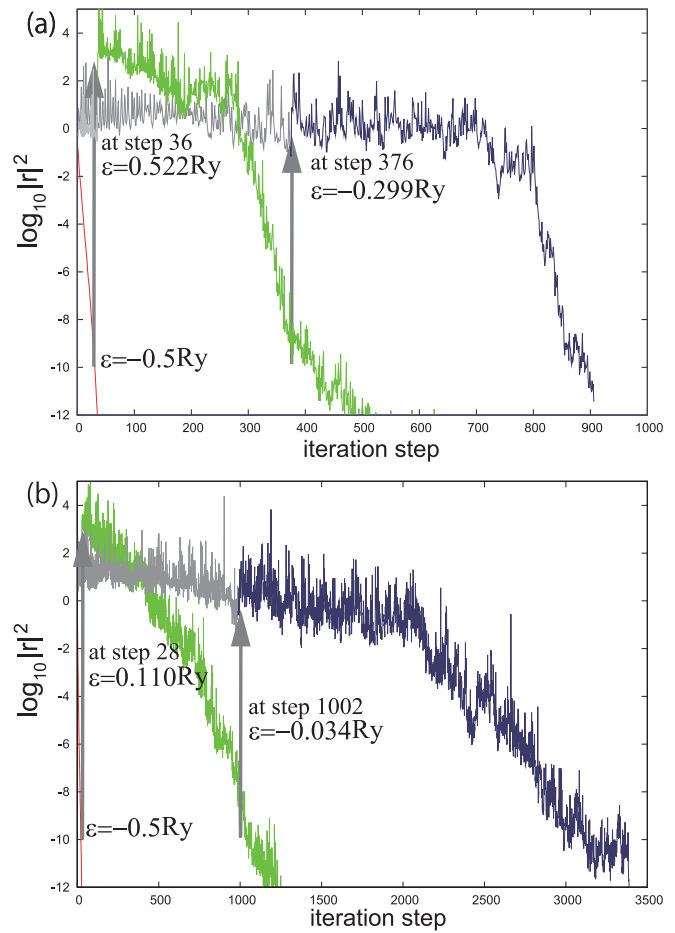}
\caption{(Color on line) Residual norm at the seed energies for a system of  Au 864 atoms   
by the NRL tight-binding Hamiltonian,~\cite{tb-NRL}  in s-orbitals (a) and d-orbitals (b). 
(a) Seed-switches to $0.0$~Ry from $-0.5$~Ry at 36-th step  and to $-0.299$~Ry at 376-th step 
in the s-orbital case.  
(b) Seed-switches to $0.110$~Ry from $-0.5$~Ry at 28-th step and to $-0.034$~Ry at 1002-nd step 
in the d-orbital case. 
Once the calculation using one seed is converged and full convergence has not been achieved, 
one should choose the next seed and continue the calculation. 
The gray lines show the residual norm by energy shift ${\bm r}_n^{\prime \sigma}$ before seed-switching.  
 }
\label{fig_GsCOCG-SeedSwitch}
\end{center}
\end{figure}

%------------------------------------------------------------------------
\subsection{Seed-switching technique}
When one chooses a seed energy in an energy range of rapid convergence, 
the spectra at majority energy points have not been converged yet 
and one should restart the calculation with a new seed energy as seen in Fig.\ref{fig_GsCOCG-ResidualNorm}a. 
The most desirable seed energy may be the one of the largest (partial) DOS 
because the convergence at these point is the most slowest. 

However, even if one chooses a starting seed energy in the highest DOS region 
and the residual norm at the seed energy reaches the convergence criterion, 
it often happens that there still remain several energy points/regions 
where the residual norm has not been small enough. 
Fortunately the shifting energy does not need any additional heavy computational task 
but several scalar manipulation as Eq.(\ref{eq(B.32)}). 
Because of this property of shifting energy, a choice of a seed energy $\sigma_s$ can be arbitrary. 
As shown in Fig.\ref{fig_GsCOCG-ResidualNorm}b, even if we start with an improper seed energy 
and switch a seed, the total iteration times for desired convergence over whole energy range is 
not very different. 
The {\it seed-switching} is very efficient technique to avoid restarting the calculation 
from the beginning with a new seed energy.~\cite{SOGABE2007,yamamoto2007b} 
One chooses a new seed energy $\sigma_s^{\rm new}$ 
and can continue the calculation without discarding the information of the previous calculation 
with the old $\sigma_s$ by using the shift property. 
Figure \ref{fig_GsCOCG-SeedSwitch} shows the behavior of the residual norms in the seed-switching process.

%%=================================== 
%%=================================== 
\section{Generalized Lanczos process and density of states}\label{GLanczos} 
The three-term recursive relation used in GsCOCG method  
leads us to the generalization of  the Lanczos method.~\cite{Haydock-1980,g-Eig-Bai,Ozaki-Terakura-2001} 
As far as we know, the generalization of the Lanczos method was presented first 
in Ref.~\onlinecite{Haydock-1980}. 
In this section, we will stress that the generalized Gram-Schmidt reorthogonalization process 
makes G-Lanczos method practically useful 
and also the use of Eqs.(\ref{eq(4.84)}) and (\ref{eq(4.87)}) gives very efficient and accurate results.

%------------------------------------
\subsection{Generalized Lanczos process}

First we define a matrix ${\cal H}$ as 
\begin{eqnarray}
{\cal H} \equiv S^{-1}H ,
\label{eq(1.10)}
\end{eqnarray}
which is not self-adjoint but still satisfies the {\it quasi-Hermitian} property in the S-product;  
\begin{eqnarray}
  ({\bm v}, {\cal H}{\bm u})_S =({\cal H}{\bm v}, {\bm u})_S .
\label{eq(2.12)}
\end{eqnarray} 
We can construct the three-term  procedure of the Lanczos process 
($n=0,\ 1 \cdots)$ as~\cite{Haydock-1980,g-Eig-Bai}
\begin{eqnarray}
{\cal H}{\bm u}^n &=& a_n{\bm u}^n +b_{n+1}{\bm u}^{n+1} +b_n{\bm u}^{n-1} , \label{eq(2.17)} 
\end{eqnarray}
where 
\begin{eqnarray}
        a_n       &=& ({\bm u}^n, {\cal H}{\bm u}^n)_S , 
                      \nonumber \\ 
        b_{n+1}^2 &=& (({\cal H}-a_n){\bm u}^n-b_n{\bm u}^{n-1},({\cal H}-a_n){\bm u}^n-b_n{\bm u}^{n-1})_S , 
                      \nonumber \\ 
    {\bm u}^{n+1} &=& \{({\cal H}-a_n){\bm u}^n-b_n{\bm u}^{n-1}\}/b_{n+1}  
                      \nonumber    
\end{eqnarray}
with conditions  $b_0=0$, $b_n\ge 0$ and then the vectors $\{{\bm u}^{m}\}$ satisfy the S-orthogonality
\begin{eqnarray}
({\bm u}^{n}, {\bm u}^{m})_S=\delta_{nm} .
\label{eq(2.21)}
\end{eqnarray}
This process we call the generalized Lanczos (G-Lanczos) process (method). 
It is well-known that the orthogonality relation is broken for larger $n$ in the Lanczos method 
and this is also the case here. 
We adopt the {\it modified Gram-Schmidt} reorthogonalization process 
in order to keep the $S$-orthogonality. 
(See the results without the {\it modified Gram-Schmidt} reorthogonalization process 
in Appendix~\ref{non-GS-method}.)

We then stop the Lanczos process up to $n=N$ and assume ${\bm u}^m=0$ ($m=N+1,N+2, \cdots$).  
This procedure constructs the Krylov subspace 
\begin{eqnarray}
K_N({\cal H}, {\bm b})=({\bm b}, {\cal H}{\bm b}, {\cal H}^2{\bm b}, \cdots,{\cal H}^N{\bm b} )   
\end{eqnarray} 
and the matrix ${\cal H}$ is transformed in this subspace to a matrix of a tridiagonal form. 

%%---------------------- Table I ------------------------------
\begin{table*}[t]
\begin{minipage}{\textwidth}
\begin{center} 

\caption{The generalized Lanczos process applied to a system of 864 atoms of fcc Au 
and comparison with that by GsCOCG. 
The Hamiltonian is the NRL tight-binding form.~\cite{tb-NRL}  
The chemical potential and the total energy are in Ry unit and $N_{\rm atom}$ is the number of atoms in the system.}
\label{tab_1b}
\begin{tabular}{cc||c|ccc||c|c}
               &                   &  \multicolumn{4}{|c||}{G-Lanczos}                                                         &\multicolumn{2}{c}{GsCOCG}\\ \hline\hline
               &                   &                               &{\small $\mu$ by Eq.(\ref{eq(4.79)})}&{\small $\mu$ by Eq.(\ref{eq(4.80)})}&{\small $\mu$ by Eq.(\ref{eq(4.81)})}& &{\small $\mu$ by Eq.(\ref{eq(4.79)})}    \\ \hline 
$\mu$          &                   &                               &  0.3010 8669     &  0.2926 6528     & 0.2870 0985      &     & 0.3006 0832 \\ 
\hline
               &Eq.(\ref{eq(4.79)})&                               &{\bf 5.5000 0001} &  5.4809 9705     & 5.4674 0401      &Eq.(\ref{eq(4.79)})& {\bf 5.5000 0000} \\
$N_{\rm tot}/2N_{\rm atom}$&Eq.(\ref{eq(4.80)})&                               &  5.5180 6152     &{\bf 5.4999 9998} & 5.4865 0739      &     & ---         \\
               &Eq.(\ref{eq(4.81)})&                               &  5.5316 2252     &  5.5135 2184     &{\bf 5.4999 9999} &     & ---         \\
\hline
               &Eq.(\ref{eq(4.88)})&                               &{\bf -0.1436 6686}&  -0.1492 6754    & -0.1531 8294     &Eq.(\ref{eq(4.88)})& {\bf -0.1436 4084}\\
$E_{\rm tot}/2N_{\rm atom}$&Eq.(\ref{eq(4.89)})&{$\rho$ by Eq.(\ref{eq(4.83)})}& -0.1311 6499     &{\bf -0.1364 7104}& -0.1403 4132     &$\rho$ by Eq.(\ref{eq(B.19)})  & -0.1251 7255 \\
               &                   &{$\rho$ by Eq.(\ref{eq(4.84)})}& -0.1316 2401     &  -0.1369 4155    &{\bf -0.1408 2023}&     &             \\
               &Eq.(\ref{eq(4.90)})&{$\pi$  by Eq.(\ref{eq(4.86)})}& -0.1311 6490     &{\bf -0.1364 7094}& -0.1403 4123     & $\pi$ by Eq.(\ref{eq(B.19-p)})& -0.1436 4084 \\
               &                   &{$\pi$  by Eq.(\ref{eq(4.87)})}& -0.1316 2391     &  -0.1369 4146    &{\bf -0.1408 2014}&     &             \\
\hline 
\end{tabular}
\end{center}
\end{minipage}
\end{table*}
%------------------------------------
%%=================================== 

Starting with a natural basis ${\bm u}^0 = {\bm e}_{j_0}$,   
one generates vectors ${\bm u}^n$ in the Krylov subspace and 
each vector corresponds to orthonormalized linear combination of atomic orbitals (LCAO);
\begin{eqnarray}
{\bm u}^m \Rightarrow  \varphi^m ({\bm r}) = \sum_j \phi_j({\bm r}) u_j^m  .
\label{eq(4.1)}
\end{eqnarray}
The normalized eigen-states in the generated Krylov subspace is denoted by 
\begin{eqnarray}
 \psi^\alpha({\bm r}) 
&=& \sum_{n=0}^{N} \varphi^n({\bm r}) Q_n^{(\alpha)} =\sum_j \phi_j({\bm r}) w_j^\alpha  ,
\label{eq(4.42)}  
\end{eqnarray}
which satisfies the Schr\"odinger equation 
\begin{eqnarray}
\sum_m{\cal H}_{nm}Q_m^{(\alpha)}=\varepsilon_\alpha Q_n^{(\alpha)} ,
\label{eq(4.42-p)}  
\end{eqnarray}
where ${\cal H}_{nm}=({\bm u}^n,{\cal H}{\bm u}^m)_S = ({\bm u}^{n})^tH{\bm u}^m$.

%- - - - - - - - - - Figure 4 GS-renormalizatio n- - - - - - - - - -%
\begin{figure}[b] 
\begin{center}
\includegraphics[width=\linewidth]{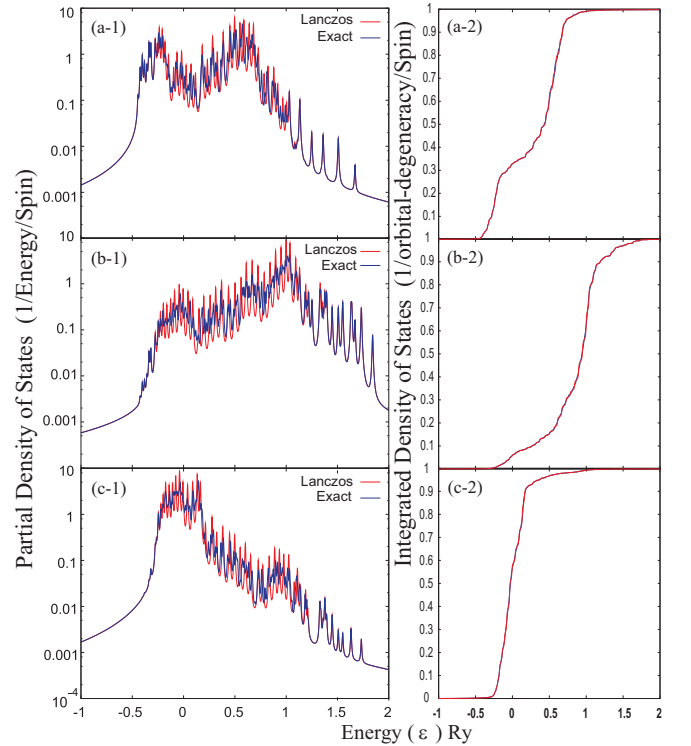}
\caption{(Color on line) Partial density of states (pDOS), normalized to unity, 
and integrated density of states (IDOS) for a system of Au 864 atoms by NRL 
tight-binding Hamiltonian.~\cite{tb-NRL}  
Comparison is between those of the generalized Lanczos method (red solid lines) and 
those of the exact ones (blue solid lines). 
(a-1) and (a-2) :pDOS and IDOS for s-orbitals. 
(b-1) and (b-2) :pDOS and IDOS for p-orbitals. 
(c-1) and (c-2) :pDOS and IDOS for d-orbitals. 
$\eta=5.0\times 10^{-3}$Ry.
}
\label{fig_e-GL_Reorthogonalization}
\end{center}
\end{figure}
%-------------------------------------------------------------------%

%------------------------------------
\subsection{Numerical test with NRL Hamiltonian for fcc Au}

Chemical potential $\mu$ can be evaluated by using Eqs.~(\ref{eq(4.79)})$\sim$(\ref{eq(4.81)}) 
in the generalized Lanczos method. 
Calculation of the Green's function uses Eq.(\ref{eq(3.8)}) having a double summation 
of atomic sites and orbitals and it consumes a long CPU time. 
On the contrary, the calculation of the density of states by Eq.~(\ref{eq(3.18)}) costs less CPU time. 
The computational efficiency will be discussed later in Section~\ref{Comparison}.
Here in this subsection, 
we show several evaluated values, the density of states, the integrated density of states 
as functions of energies for a system of gold 864 atoms of fcc structure 
described by the tight-binding Hamiltonian constructed by Mehl and Papaconstantpolous.~\cite{tb-NRL}

Several evaluated values and consistency between them are summarized in Table \ref{tab_1b}.
The parameters in the generalized Lanczos method are 
$N=50$, the convergence criterion $\delta=10^{-6}$Ry in the inner CG process of ${\bm r}^\prime=S^{-1}{\bm r}$. 
In GsCOCG, the imaginary small energy $\eta=10^{-3}$Ry, 
the total number of the energy integration mesh-points is 3,000, and 
the convergence criterion $\delta=10^{-6}$Ry in the inner CG and outer iteration procedures. 
The difference of the calculated total energy is of the order of $10^{-2} \sim 10^{-3}$Ry.
The scale of the band energy is 1~Ry and the relative error may be of $10^{-3}$. 
The bold numbers in Table \ref{tab_1b} are a set of consistent values in each case 
(for each equation determining the chemical potential) and 
the combination of Eqs.~(\ref{eq(4.83)}) and (\ref{eq(4.86)}), 
and that of Eqs.~(\ref{eq(4.84)}) and (\ref{eq(4.87)}) are consistent pairs of data.
This consistency between the density matrix and energy density matrix is 
crucial for consistency between the total energy and force 
(Appendix~\ref{EF-consistency}).

In GsCOCG method, calculation of Eqs.~(\ref{eq(B.19)}) and (\ref{eq(B.19-p)})  needs  
$\varepsilon$-integration  of $\rho_{ij}(\varepsilon)$ 
and the numerical integration causes a certain error 
in the integration of the tail of the spectra. 
Once we reduce the value of $\eta$ and increase the $\varepsilon$-points of the integration, 
this discrepancy can be reduced. 
In the generalized Lanczos method, 
the combination of Eqs.~(\ref{eq(4.81)}), (\ref{eq(4.84)}) and (\ref{eq(4.87)}) is the best scheme,
since we do not use the numerical energy integral.

Figure \ref{fig_e-GL_Reorthogonalization} shows the partial density of states and integrated density of states 
for s-, p-, and d-orbitals. 
Those by the generalized Lanczos method are compared with exact results. 
The peak positions and the tails overlap excellently and 
we can conclude that the generalized Lanczos method can be powerful and convenient tool. 
One must notice that the exact results are more smooth in the central energy region, 
since the correct density of states is more dense in this energy region. 
In other words, the exact results reserve entire profile with $864 \times 9$ peaks but, 
on the contrary, those by the generalized Lanczos method reserve the profiles with 51($=N+1$) peaks  
or the density of states is expressed as a polynomial function of energy of the order 51 
in the present calculation.  

%----------------------- Table II ---------------------------------
\begin{table*}[tb]
\begin{minipage}{\textwidth}
\begin{center}
{\small 
\caption{The generalized Arnoldi process applied to a system of 864 atoms of fcc Au and  comparison with that by GsCOCG. 
The Hamiltonian is the NRL tight-binding form.~\cite{tb-NRL}  
The chemical potential and the total energy are in Ry unit and $N_{\rm atom}$ is the number of atoms in the system.}
\label{tab_1d}
\begin{tabular}{cc||c|ccc||c|c}
               &                   &  \multicolumn{4}{|c||}{G-Arnoldi}                                                         &\multicolumn{2}{c}{GsCOCG}\\ \hline\hline
               &                   &                             &{\small $\mu$ by Eq.(\ref{eq(4.79)})}&{\small $\mu$ by Eq.(\ref{eq(4.80)})}&{\small $\mu$ by Eq.(\ref{eq(4.81)})}& &{\small $\mu$ by Eq.(\ref{eq(4.79)})} \\ \hline  
$\mu$          &                   &                             &  0.3039 0679     &   0.2928 7198    &  0.2864 6136    &                & 0.3006 0832 \\  \hline
               &Eq.(\ref{eq(4.79)})&                             &{\bf  5.5000 0002}&   5.4797 1790    &  5.4658 1773    &Eq.(\ref{eq(4.79)})&{\bf 5.5000 0000} \\
$N_{\rm tot}/2N_{\rm atom}$&Eq.(\ref{eq(4.80)})&                             &  5.5176 0887     &{\bf 5.5000 0001} &  5.4865 0981    &                &---          \\
               &Eq.(\ref{eq(4.81)})&                             &  5.5311 6653     &   5.5135 1949    &{\bf 5.4999 9999}&                &---          \\\hline
               &Eq.(\ref{eq(4.88)})&                             &{\bf -0.1423 0073}&   -0.1482 6563   & -0.1522 3179     &Eq.(\ref{eq(4.88)})&{\bf -0.1436 4084}\\
$E_{\rm tot}/2N_{\rm atom}$&Eq.(\ref{eq(4.89)})&$\rho$ by Eq.(\ref{eq(4.83)})& -0.1300 1076     &{\bf -0.1351 4434}& -0.1389 5498     &$\rho$ by Eq.(\ref{eq(B.19)}) &-0.1251 7255 \\
               &                   &$\rho$ by Eq.(\ref{eq(4.84)})& -0.1304 6635     & -0.1356 1105     &{\bf -0.1394 2996}&               &             \\
               &Eq.(\ref{eq(4.90)})&$\pi$  by Eq.(\ref{eq(4.86)})& -0.1300 1076     &{\bf -0.1351 4434}& -0.1389 5498     &$\pi$ by Eq.(\ref{eq(B.19-p)})&-0.1436 4084 \\
               &                   &$\pi$  by Eq.(\ref{eq(4.87)})& -0.1304 6635     & -0.1356 1105     &{\bf -0.1394 2996}&               &             \\
\hline 
\end{tabular}
}
\end{center}
\end{minipage}
\end{table*}
%- - - - - - - - - - Table - - - - - - - - - -%
%%=================================== 
%%=================================== 
\section{Generalized Arnoldi process and density of states}\label{GArnoldi}

%------------------------------------
\subsection{Generalized Arnoldi process}
We can construct the Krylov subspace, starting with a natural basis ${\bm u}^0 = {\bm e}_{j_0}$, 
by using the Hamiltonian matrix $H$, as~\cite{g-Eig-Bai,Hoshi-etal-2010}
\begin{eqnarray}
 {\bm l}^{n+1} &=& H{\bm u}^n                                                      \label{eq(D.3)} \\
 {\bm k}^{n+1} &=& {\bm l}^{n+1}-\sum_{m=0}^n {\bm u}^m ({\bm u}^m, {\bm l}^n)_S   \label{eq(D.4)} \\
 {\bm u}^{n+1} &=& \frac{{\bm k}^{n+1}}{{({\bm k}^{n+1},{\bm k}^{n+1})_S}^{1/2}}.   \label{eq(D.5)}
\end{eqnarray}
This is the Arnoldi process and we call it the generalized Arnoldi (G-Arnoldi) method. 
The generalized Arnoldi method generates the Krylov subspace 
\begin{eqnarray}
 K_{N+1}(H;{\bm b}) 
&=& {\rm Span}\{{\bm b},H{\bm b}, H^2{\bm b}, \cdots , H^N{\bm b}\} \nonumber \\
&=& {\rm Span}\{{\bm u}^0, {\bm u}^1, {\bm u}^2, \cdots {\bm u}^N\} .
\label{eq(D.1)}
\end{eqnarray}
The generated vector ${\bm u}^m$ corresponds to orthonormalized LCAO 
\begin{eqnarray}
\varphi^m ({\bm r}) = \sum_j \phi_j({\bm r}) u_j^m  
\end{eqnarray}
as in G-Lanczos method.
An eigen-function 
$\psi^\alpha({\bm r}) = \sum_n \varphi^n({\bm r}) Q_n^{(\alpha)}$
satisfies the Schr\"odinger equation 
\begin{eqnarray}
\sum_m {\tilde H}_{nm} Q_m^{(\alpha)} =\varepsilon_\alpha Q_n^{(\alpha)}
\label{eq(D.15)}
\end{eqnarray}
where ${\tilde H}_{mn} =  \langle \varphi^m|{\hat H}|\varphi^n \rangle  = ({\bm u}^m)^{\rm t}H{\bm u}^n$ 
and ${\tilde H}$ is an upper Hessenberg matrix.  
We can say that this procedure is a kind of generalization of the subspace diagonalization 
of the Krylov subspace developed before.~\cite{TAKAYAMA2004}

%- - - - - - - - - - Figure 5 Arnoldi - - - - - - - - - -%
\begin{figure}[th]
\begin{center}
\includegraphics[width=\linewidth]{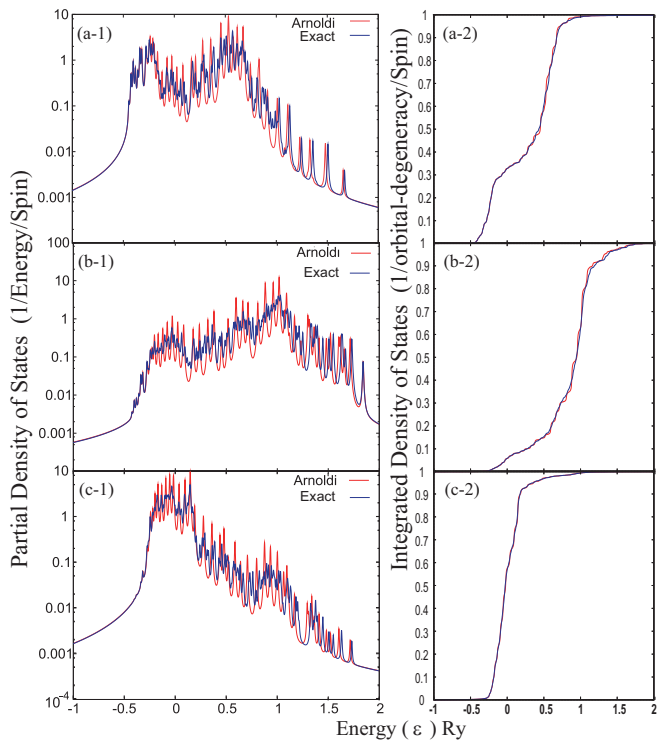}
\caption{(Color on line) Partial density of states, normalized to unity, and integrated density of states. 
Comparison is between those of the generalized Arnoldi method (red solid lines) and 
those of the exact ones (blue solid lines), for a system of Au 864 atoms by NRL 
tight-binding Hamiltonian.~\cite{tb-NRL}  
(a-1) and (a-2) :pDOS and IDOS for s-orbitals. 
(b-1) and (b-2) :pDOS and IDOS for p-orbitals. 
(c-1) and (c-2) :pDOS and IDOS for d-orbitals. 
$\eta=5.0\times 10^{-3}$Ry.
}
\label{fig_Arnoldi-Lanczos}
\end{center}
\end{figure}
%- - - - - - - - - - Figure - - - - - - - - - -%

%------------------------------------
\subsection{Numerical test with NRL Hamiltonian for fcc Au}

We summarize, in Table \ref{tab_1d}, several evaluated values and consistency between them, 
in comparison with the results of GsCOCG. 
The system is of 864 atoms of fcc Au by NRL tight-binding Hamiltonian.~\cite{tb-NRL}
The bold numbers in the table are a set of consistent results in each calculation of the 
chemical potential. 
$N=50$ for the generalized Arnoldi process. 
Data of GsCOCG are the same as in Table \ref{tab_1b}. 
The calculated values of the total energies agree with those by the generalized Lanczos method 
and the overall difference is less than 1\% as shown in Tables \ref{tab_1b}.

Figure~\ref{fig_Arnoldi-Lanczos} shows the partial density of states and the integrated density of states 
as functions of energies.  
The peak positions are deviated slightly from those by the exact calculation, 
which one could make smaller with increasing a dimension $N$ of the Krylov subspace. 

%%=================================== 
%%=================================== 
\section{Comparison among GsCOCG, G-Lanczos method and G-Arnoldi method}\label{Comparison}
%------------------------------------
\subsection{Convergence}
The dimension of the Krylov subspace in GsCOCG, G-Lanczos or G-Arnoldi methods 
equals to $N+1$ where $N$ is the maximum iteration step. 
GsCOCG method is very accurate method if one achieves the iteration to have enough small residual norm 
(e.g. $\delta=10^{-6}$Ry). 
In Fig.\ref{fig_GsCOCG-ResidualNorm},  we have shown the convergence behavior of the residual norms 
with different seed energies and until much smaller convergence region. 
One should use the same iteration criteria 
both in the inner CG and outer procedures in GsCOCG method.  
It sometimes happens that the resultant DOS shows an un-physical behavior, e.g. negative values of DOS,  
if one stops the iteration steps before enough convergence in GsCOCG method. 
On the other hand, G-Lanczos and G-Arnoldi methods never give such un-physical DOS 
even if one stops at small iteration step because of the expression of Eq.(\ref{eq(3.8)}). 
Furthermore, the first $N$ moments are preserved correctly in the energy spectra of G-Lanczos method. 
In the spectrum of the present model by the exact calculation, 
we observe about forty prominent peaks 
and then we use $N=50$ in the calculations of G-Lanczos and G-Arnoldi procedures. 
This is presumably the reason why the peak positions and detailed profiles in 
the spectra of G-Lanczos method show excellent agreement with those of the exact calculation. 
G-Lanczos method needs the Gram-Schmidt reorthogonalization and also it is necessary 
to have enough convergence in the inner CG procedure. 
G-Arnoldi method does not need such reorthogonalization procedure since one solve 
the eigen-value problem in that subspace.

In condensed matters, the width of the valence and/or conduction bands $W$ may be of the order of 1Ry. 
Then, when the number of atoms is $N_{\rm atom}$, we can estimate the separation of each energy level 
as of the order of $W/(9 \times N_{\rm atom})$. 
Presumably 10\% of this separation would be an enough accuracy in the energy scale.  
In our present case, with $N_{\rm atom} \approx1,000$, the convergence criterion can be chosen as 
$0.1\times W/(9 \times N_{\rm atom}) \approx 0.1 \times 1/9,000 \approx 10^{-5}$Ry. 
We also observed that the maximum iteration steps are almost the same 
for the convergence criterion $\delta=10^{-5}$ and $\delta=10^{-6}$ in GsCOCG and G-Lanczos methods. 
This is the reason why we choose $\delta =10^{-6}$Ry. 

%%---------------------- Table III ------------------------------
\begin{table}[t]
\begin{center} 
\caption{Chemical potential and the total energy (in Ry unit) of systems of 864 and 256 atoms 
of fcc Au~\cite{tb-NRL}  by three different method.
The values of $\rho$ in the G-Lanczos and G-Arnoldi methods are evaluated by  Eq.(\ref{eq(4.84)}).
}  
\label{tab_III}
\begin{tabular}{cc|c c c}
      & {G-Lanczos}                  &  (864 atoms) & \ \ & (256 atoms)    \\ \hline
$\mu$ & Eq.(\ref{eq(4.80)})          &  0.2926 6528 &     & 0.2875 4133    \\ 
$E_{\rm tot}/2N_{\rm atom}$ &Eq.(\ref{eq(4.89)}) & -0.1364 7104 &     &-0.1358 7639    \\ \hline \hline
      & {G-Arnoldi}                  &  (864 atoms) & \ \ & (256 atoms)    \\ \hline
$\mu$ & Eq.(\ref{eq(4.80)})          &  0.2928 7198 &     & 0.2886 2149    \\ 
$E_{\rm tot}/2N_{\rm atom}$ &Eq.(\ref{eq(4.89)}) & -0.1351 4434 &     &-0.1346 6195    \\ \hline \hline
      & {GsCOCG}                     &  (864 atoms) & \ \ & (256 atoms)    \\ \hline
$\mu$ & Eq.(\ref{eq(4.79)})          & 0.3006 0832  &     & 0.2947 1535    \\ 
$E_{\rm tot}/2N_{\rm atom}$ &Eq.(\ref{eq(4.88)}) & -0.1436 4084 &     &-0.1433 1925    \\ \hline 

\end{tabular}
\end{center}
\end{table}
%%-------------------------------------------------------------

We compare the results of the chemical potential and the total band energy 
of systems of 256 and 864 atoms in Table \ref{tab_III}. 
The difference of the  chemical potential $\mu$ is of the order of $10^{-3}$Ry, and
that of the total band energy $E_{\rm tot}$ is of the order of $10^{-4}$Ry. 
The level separation in 256 atom system can be estimated as $1.0/(9\times 256)\approx 4\times 10^{-4}$Ry 
and that in 864 atom system $10^{-4}$Ry. 
The difference in the chemical potential of two systems of different sizes 
is due to the difference of total number of levels 
which changes the value of the chemical potential sensitively. 
On the other hand, the difference of $E_{\rm tot}$ is just the quantity related to the overall spectrum
and we can see an excellent convergence of the results of $N=50$.

\begin{figure}[bth]
\begin{center}
\includegraphics[width=\linewidth]{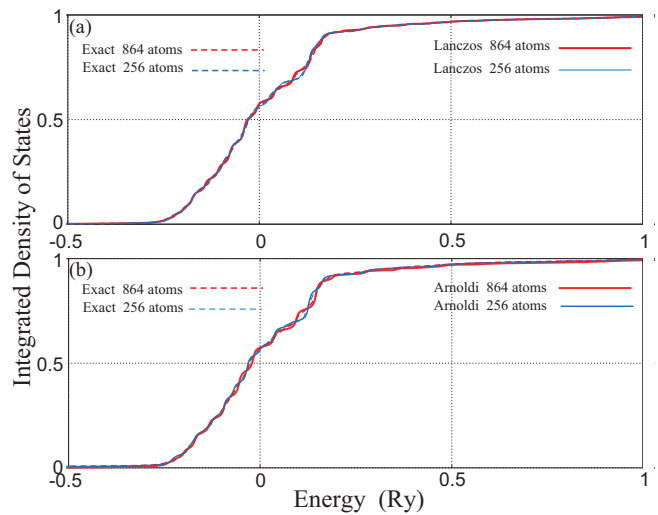}
\caption{(Color on line) Comparison of IDOS of d-orbitals for systems of Au 256 and 864 atoms by NRL 
tight-binding Hamiltonian.~\cite{tb-NRL} 
(a) G-Lanczos (solid line) and exact calculation (chain line) for 864 atoms (red) and 256 (blue) atoms. 
(b) G-Arnoldi (solid line) and exact calculation (chain line) for 864 atoms (red) and 256 (blue) atoms. 
The agreement between the results by G-Lanczos or G-Arnoldi methods and those by the exact calculation 
is excellent. 
}
\label{fig_Enlarged-Scaling}
\end{center}
\end{figure}

Figure \ref{fig_Enlarged-Scaling} shows the actual convergence behavior of IDOS 
by G-Lanczos and G-Arnoldi method. 
The agreement between the results by G-Lanczos or G-Arnoldi methods and those by the exact calculation 
is excellent both for systems of 864 atoms and that of 256 if we adopt $N=50$. 
The most apparent difference appears in the IDOS curves of 
exact calculation of systems of 256 atoms and 865 atoms in the mid energy region,  
and the calculated results by our present methods present this difference with complete fidelity. 

%- - - - - - - - - - Figure 6scaling - - - - - - - - - -%
%- - - - - - - - - - Figure - - - - - - - - - -%

%- - - - - - - - - - Table IV - - - - - - - - - -%
\begin{table}[hbt]
\caption{CPU times by using a standard single CPU workstation, 
for (a) a system of gold 256 atoms (Au256) and (b) gold 864 atoms (Au864) 
by NRL tight-binding Hamiltonian.~\cite{tb-NRL}  
}
\label{tab_0ab}
\begin{center}
\begin{tabular}{l|c|cc}
 (a) Au 256&         &   CPU-times (s)            &        \\ \hline
G-Arnoldi  &main part&  0.52 \ \ \ \ 1.11         & ---    \\
           &total    &  1.59 \ \ \ \ 3.20         & ---    \\ \hline
G-Lanczos  &inner CG &  2.04 \ \ \ \ 3.89         & ---    \\
           &main part&  2.89 \ \ \ \ 5.60         & ---    \\
           &total    &  3.92 \ \ \ \ 7.62         & ---    \\\hline
GsCOCG     &seed     &    ---                     &13.67   \\
           &shifted  &    ---                     &7.53    \\
           &total    &    ---                     &21.20   \\\hline 
Exact      &         &    ---                     &57.60   \\\hline 
\hline 
           &         &                            &        \\
 (b) Au 864&         &   CPU-times (s)            &        \\ \hline
G-Arnoldi  &main part&  1.94 \ \ \ \ 3.76         & ---    \\
           &total    &  3.00 \ \ \ \ 6.00         & ---    \\ \hline
G-Lanczos  &inner CG &  7.93 \ \ \ \ 15.13        & ---    \\
           &main part& 10.87 \ \ \ \ 21.13        & ---    \\
           &total    & 11.93 \ \ \ \ 23.31        & ---    \\\hline
GsCOCG     &seed     &    ---                     & 140.89 \\
           &shifted  &    ---                     & 108.36 \\
           &total    &    ---                     & 249.15 \\\hline 
Exact      &         &    ---                     &2111.51 \\\hline

\end{tabular}
\end{center}
\end{table}
%- - - - - - - - - - Table - - - - - - - - - -%
%------------------------------------
\subsection{CPU times}
We summarize, in Tables \ref{tab_0ab}, the CPU times (by using single CPU of the standard workstation) 
for (s-orbitals) $D_{ii}$ of the generalized Lanczos and the generalized Arnoldi methods with Eq.(\ref{eq(3.18)}), 
    that of GsCOCG with Eq.(\ref{eq(B.20)}), 
and that of the exact diagonalization method  
for the NRL Hamiltonian of fcc Au system of 256 and 864 atoms.~\cite{tb-NRL}  
The total number of orbitals equals to nine times of the total number of atoms (1 s, 3 p's and 5 d's). 
We use, in the inner CG process of the generalized Lanczos (G-Lanczos) method, 
the convergence criterion $\delta=10^{-6}$Ry. 
Two numbers in the row of the CPU time are referred to those of $N=50$ and $N=100$, respectively, for 
G-Lanczos and G-Arnoldi methods, though the results of $N=100$ almost coincide with 
those of $N=50$.  
For GsCOCG (with shifted 3,000 energy points), 
the data shown here are those of $\delta=10^{-6}$Ry both for in the inner and outer iteration processes. 
The repeated time of the inner CG process ($S^{-1}{\bf x}$ part) in GsCOCG and G-Lanczos method 
is $10 \sim 11$.  (Repeated time of $25 \sim 27$ is needed for $\delta=10^{-18}$Ry.)

The system size dependence of the CPU time is 
linear for the generalized Arnoldi (G-Arnoldi) method and the generalized Lanczos (G-Lanczos) method, 
bilinear for GsCOCG method and cubic for the exact calculation. 
The generalized Arnoldi method is extremely efficient in electronic structure calculations 
of extra-large systems with several hundred thousands atoms.

%------------------------------------
\subsection{Applicability to large-scale electronic structure calculations and MD simulations}
In the exact calculation and GsCOCG method, calculations of physical properties, such as 
the density matrix, the energy density matrix, and chemical potential, 
require the numerical integration as in Eqs.~(\ref{eq(4.79)}),   
 (\ref{eq(4.88)}), (\ref{eq(B.19)}) and (\ref{eq(B.19-p)}). 
Therefore, in order to keep high accuracy, the integration needs fine energy mesh points. 
On the other hand, the generalized Lanczos or the generalized Arnoldi methods use the 
simple summation of the eigen-states in the mapped subspace in 
Eqs.~(\ref{eq(4.80)}), (\ref{eq(4.81)}), and (\ref{eq(4.83)}) $\sim$ (\ref{eq(4.87)}). 
These two methods do not consume the CPU time and give stable values of 
the density matrix and the energy density matrix.

The CPU times per one MD-step are, for the present models, a few seconds by 
the generalized Lanczos method and the generalized Arnoldi method. 
From the above comparison among various viewpoints, 
we can conclude that the generalized  Lanczos method or the generalized  Arnoldi method 
are very suitable to large-scale electronic structure calculations and 
MD simulation of several tens of thousands atoms and a long MD-steps. 
On the other hand, GsCOCG method can give excellently rigorous results with more CPU times 
and may be applicable to problems of a fixed atomic configuration (but not for the MD simulation).

GsCOCG method is based on the three-term recursive equations and 
we need store three generated vectors at each recursive process. 
Of course, when the size of the Hamiltonian and overlap matrices are extremely large 
and the memory size becomes a serious obstacle, 
(though much smaller consumption than the exact diagonalization method, ) 
we should invent other method of much faster convergence and smaller cost of memory size.

The convergence criterion $\delta=10^{-5}$ might corresponds to the range of neighboring 1,000 atoms, 
as already discussed, and we do not observe any clear difference between results by 
the present methods and the exact calculations. 
Even when we should discuss some physics of nano-scale systems, the electronic structure 
is determined by some nearby surroundings. 
This idea we call  {\it near-sitedness}.~\cite{Kohn1996} 
Even when we have to deal with a much larger systems, 
we can use smaller interaction range than the system size due to the near-sitedness. 
Presumably more serious problem of the system size 
in some specific problems, 
for examples, {\it entire} calculation of nano-devise or the electron-strain field interaction 
such as fracture propagation~\cite{Si-fracture-1,Si-fracture-2} and 
dislocation.~\cite{Miyata2001}

%====================================
%%=================================== 
\section{Conclusion}\label{Conclusion}

We have derived several efficient and accurate algebraic methods to calculate the 
Green's functions, total/partial density of states and total band energy  
in case of non-orthogonal atomic orbitals. 
The method is very general. 

We have investigated the accuracy and efficiency by showing numerical data 
with different numerical procedures. 
GsCOCG is very accurate with less consumption than the exact diagonalization 
but may not be appropriate for long MD-step simulations. 
The generalized Lanczos method becomes applicable to actual large systems with 
the modified Gram-Schmidt reorthogonalization to maintain the orthogonality of 
generated basis vectors. 
Then, the generalized Arnoldi method and the generalized Lanczos method are accurate and efficient, 
and their CPU times depend linearly upon the system size.  
Therefore, these two methods would be the most suitable to 
the large-scale electronic structure calculations and MD simulations. 
A crucial point we should point out finally is the fact that 
G-Lanczos and G-Arnoldi methods do not adopt any numerical integration in energy 
which leads additional numerical error.

%%=================================== 
%==================================== 
\section*{Acknowledgments}
One of authors (T. Fujiwara) expresses sincere thanks to TOYOTA Motor Corporation for the financial support. 
Numerical calculation was partly carried out 
using the supercomputer facilities of
the Institute for Solid State Physics, University of Tokyo.
Our research progress of large-scale systems and other information can be found 
on the WEB page of ELSES (Extra-Large Scale Electronic Structure calculation) Consortium; 
http://www.elses.jp.

%- - - - - - - - - - Figure 7  No GS-renormalization (Appendix) - - - - - - - - - -%
\begin{figure}[ht]
\begin{center}
\includegraphics[width=\linewidth]{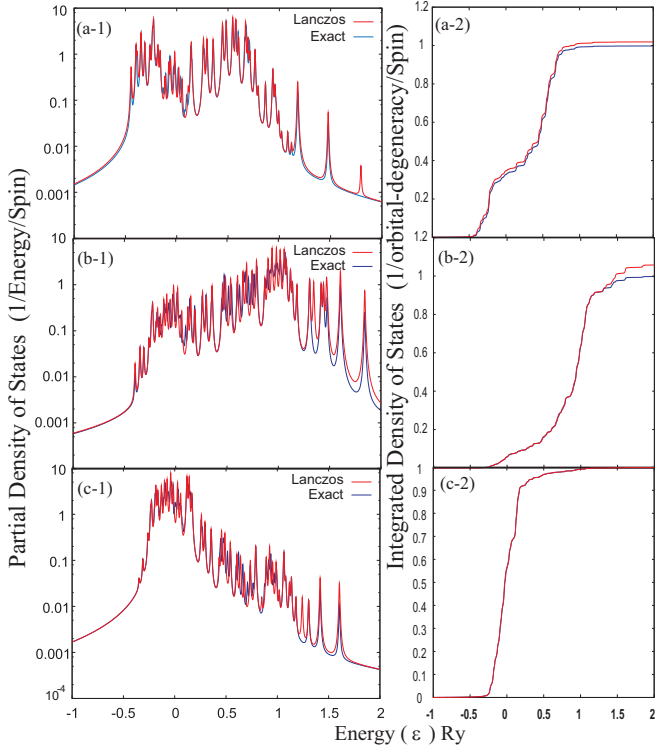}
\caption{(Color on line) Partial density of states and integrated ones without
the modified Gram-Schmidt reorthogonalization 
of the generalized Lanczos method (red solid lines) and those by the exact ones 
(blue solid lines) for a system of Au 256 atoms by NRL tight-binding Hamiltonian.~\cite{tb-NRL}
 $N=50$. 
(a-1) and (a-2) :pDOS and IDOS for s-orbitals. 
(b-1) and (b-2) :pDOS and IDOS for p-orbitals. 
(c-1) and (c-2) :pDOS and IDOS for d-orbitals. 
One should notice that the normalization of the integrated density of states is broken 
without reorthogonalization procedure and that some ``ghost" peaks appear due to 
incorrect mixing of states. 
$\eta=5.0\times 10^{-3}$Ry. 
}
\label{fig_e-GL_NoReorthogonalization}
\end{center}
\end{figure}
%- - - - - - - - - - Figure - - - - - - - - - -%

%%===================================
%%===================================  
\appendix
%------------------------------------
\section{Re-orthogonalization by modified Gram-Schmidt method}\label{non-GS-method}

The three-term recursive relation in the generalized Lanczos method 
guarantees  theoretically the automatic S-orthogonalization. 
However, the orthogonality is broken in the numerical calculation procedure. 
This problem causes several troubles such as the existence of constant background of error 
in the spectrum,~\cite{TAKAYAMA2006}  
appearance of ``ghost" structure in spectrum due to erroneous mixing of states and 
a broken normalization of the partial density of states.

Figure~\ref{fig_e-GL_NoReorthogonalization} shows the examples of this 
broken orthonormality, 
in a system of 256 atoms of fcc Au by using the NRL Hamiltonian. 
One can see the ``ghost" peaks (e.g. at $\varepsilon\simeq 1.8$Ry in (a-1), 
at $\varepsilon\simeq 1.25\sim 1.3$Ry in (b-1) and (c-1)) 
and broken normalization (e.g. in (a-2) and (b-2)).  
These problems are solved by the re-orthogonalization 
with the modified Gram-Schmidt method.

%------------------------------------
\section{Consistency between the total energy minimum and a vanishing force}\label{EF-consistency}

We should construct our eigen-states in a small subspace and 
a certain numerical error is unavoidable in evaluated total energy and force. 
Even in that case, the consistency between the total band energy minimization and 
an vanishing atomic force is the most important 
in the electronic structure calculation in equilibrium atom configuration. 
In the framework of the tight-binding model, 
the force (due to band energy)  acting on an atom $I$ is evaluated by a formula 
\begin{eqnarray}
{\bm F}_I= -2 \sum_{ij} \Big(\rho_{ij}\frac{\partial H_{ij}}{\partial {\bm R}_I}-\pi_{ij} \frac{\partial S_{ij}}{\partial {\bm R}_I}\Big) ,
\label{eq(4.add-4)}
\end{eqnarray}
which can be rewritten, 
with only an assumption of the eigen-state property  Eq.(\ref{eq(1.5)}) in the mapped subspace, 
as 
\begin{eqnarray}
 {\bm F}_I
 &=&-2\frac{\partial}{\partial {\bm R}_I} \Big\{\sum_{ij}(\rho_{ij}H_{ij})-\sum_{ij}(\pi_{ij}S_{ij})\Big\} \nonumber \\
  && -\frac{\partial}{\partial {\bm R}_I}\sum_\alpha f(\varepsilon_\alpha) \varepsilon_\alpha.
\label{eq(4.add-6)}
\end{eqnarray}
Therefore, calculated atomic and electronic configuration of the minimum total energy is consistent 
with that of vanishing atomic force, 
if the identity $\sum_{ij}(\rho_{ij}H_{ij})=\sum_{ij}(\pi_{ij}S_{ij})$ is satisfied always 
in any atomic configuration. 
It should be noticed that the above equality is satisfied in the mapped subspace 
as described in \ref{ChemPot-BandEne}. 
It is important in actual calculating procedure that we should use 
the consistent pair of equations as Eqs.~(\ref{eq(4.83)}) and (\ref{eq(4.86)}) or 
Eqs.~(\ref{eq(4.84)}) and (\ref{eq(4.87)}).

%%=================================== 
%%=================================== 
% --------------- Bibliography ------

% --------------- Bibliography ------

\end{document}